\documentstyle[11pt,paspconf,epsf]{article}

\def\BP{Ballesteros-Paredes}
\def\ekin{{\cal E}_{\rm kin}}
\def\gamef{\gamma_{\rm e}}
\def\'#1{\ifx#1i{\accent"13\i}\else{\accent"13#1}\fi}
\def\Peq{P_{\rm eq}}
\def\rhot{\rho_{\rm t}}
\def\VS{V\'azquez-Semadeni}
\def\xruu{{\cal T_{\rm kin}}}

\def\ltsima{$\; \buildrel < \over \sim \;$}    % Use in text mode
\def\lesssim{\lower.5ex\hbox{\ltsima}}           % Use in math mode
\def\gtsima{$\; \buildrel > \over \sim \;$}    % Use in text mode
\def\gtrsim{\lower.5ex\hbox{\gtsima}}           % Use in math mode

\begin{document}

\title{Interstellar Turbulence, Cloud Formation and Pressure Balance}
\author{Enrique V\'azquez-Semadeni}
\affil{Instituto de Astronom\'\i a, UNAM, Apdo.\ Postal 70-264,
M\'exico 04510, D.F., MEXICO}

\begin{abstract}
We discuss HD and MHD compressible turbulence as a cloud-forming and
cloud-structuring mechanism in the ISM. Results from a numerical model
of the turbulent ISM at large scales suggest that the phase-like
appearance of the medium, the typical values of the densities
and magnetic field strengths in the intercloud medium, as
well as Larson's velocity dispersion-size scaling relation in clouds may be
understood as consequences of the interstellar turbulence. However,
the density-size relation appears to only hold for the densest
simulated clouds,
there existing a large population of small, low-density clouds, which,
on the other hand, are hardest to observe. We then discuss several tests and
implications of a fully dynamical picture of interstellar clouds. The
results imply that clouds are transient, constantly being formed, distorted
and disrupted by the turbulent velocity field, with a fraction of
these fluctuations undergoing gravitational collapse. Simulated line
profiles and estimated cloud lifetimes are consistent with
observational data. In this scenario, we suggest it is quite unlikely
that quasi-hydrostatic structures on any scale can form, and that the
near pressure balance between clouds and the intercloud medium is an
incidental consequence of the density field driven by the turbulence
and in the presence of appropriate cooling,
rather than a driving or confining mechanism.
\end{abstract}


\section{Introduction}

One of the main features of turbulence is its multi-scale nature
(e.g., Scalo 1987; Lesieur 1990). In particular, in the
interstellar medium (ISM),
relevant scale sizes span nearly 5 orders of magnitude, from the size of
the largest complexes or ``superclouds'' ($\sim 1$ kpc) to that of dense
cores in molecular clouds (a few $\times 0.01$ pc), with densities
respectively ranging from $\sim 0.1$ cm$^{-3}$ to $ \gtrsim 10^6$
cm$^{-3}$. Moreover, in the diffuse gas itself, even smaller scales,
down to sizes several $\times 10^2$ km are active, although at small
densities. Therefore, in a unified turbulent picture of the ISM,
it is natural to expect that turbulence can intervene in
the process of cloud formation (Hunter 1979; Hunter \& Fleck 1982;
Elmegreen 1993; \VS, Passot \& Pouquet 1995, 1996) through modes larger
than the clouds 
themselves, as well as in providing cloud support and determining the
cloud properties, through modes smaller than the clouds
(Chandrasekhar 1951; Bonazzola et al.\ 1987; L\'eorat et al.\ 1990;
\VS\ \& Gazol 1995). Moreover, 
another essential feature of turbulence is that all these scales
interact nonlinearly, so that coupling is expected to exist between
the large-scale cloud-forming modes and the small-scale cloud
properties.

In this chapter we adopt the above viewpoint as a framework for
presenting some of the most relevant results we have learned from
two-dimensional (2D) numerical simulations of the turbulent ISM in a
unified and coherent 
fashion, as it relates to the problems of cloud formation, the
phase-like structure of the ISM and the
topology of the magnetic and density fields, as well as internal cloud
properties, such as their virialization and scaling relations (\S\
\ref{2D}). Then, in  \S\ \ref{turb_clouds} we discuss further tests
and implications of a fully dynamical picture of interstellar clouds
and their sub-structure. \S\ \ref{corr_vir} discusses the correlation
between the field variables, and the fact that fluid velocities are
large at cloud boundaries in the simulations. \S\ \ref{hydrostatic?}
discusses the feasibility of forming hydrostatic structures within a
turbulent medium, and \S\ \ref{comp_obs} compares simulated line
profiles and estimated cloud lifetimes with the corresponding
observational data, showing there is good agreement. Then, \S\
\ref{appl_molec} discusses whether the results of the simulations can
be considered applicable to molecular gas. Finally, \S\
\ref{conclusions} presents a summary and some concluding remarks.

\section{Cloud Formation and Properties in the Turbulent ISM} \label{2D}

In a series of recent papers (\VS\ et al.\ 1995 (Paper I), 1996 (Paper
III); Passot, \VS\ \& Pouquet 1995 (Paper II)), we have
presented two-dimensional (2D) numerical simulations of
turbulence in the ISM on the Galactic plane, including self-gravity,
magnetic fields, simple parametrizations 
of standard cooling functions for atomic and ionized gas (Dalgarno \&
McCray 1972; Raymond, Cox \& Smith 1976) as fitted by Chiang \& Bregman
(1988), diffuse 
heating mimicking that of background UV radiation and cosmic rays,
rotation, and a simple prescription for star formation (SF)
which represents massive-star ionization heating by turning on a local
source of heat wherever the density exceeds a threshold
$\rhot$. Supernovae are now being included (Ga\-zol \&
Passot 1998; see also Korpi, this 
Conference, for analogous simulations in 3D). The
simulations follow the evolution of a 1 kpc$^2$ region of the ISM at
the solar Galactocentric distance over $\sim 10^8$ yr and are started
with Gaussian fluctuations with random 
phases in all variables. The initial fluctuations in the velocity
field produce shocks which trigger star formation which, in turn, feeds
back on the turbulence, and a self-sustaining cycle is
maintained. These simulations have been able to reproduce a number of
important properties of the ISM, suggesting that the processes
included are indeed relevant in the actual ISM. Some interesting
predictions have also resulted.

\subsection{Effective Polytropic Behavior and Phase-Like Structure}
\label{phases}

One of the earliest results of the simulations is a consequence of the
rapid thermal rates (Spitzer \& Savedoff 1950), faster than the dynamical
timescales by factors of 10--$10^4$ in the simulations (Paper
I). Thus, the gas is
essentially always in thermal equilibrium, except in star-forming
regions, and an effective
polytropic exponent $\gamef$ (Elmegreen 1991) can be calculated,
which results from 
the condition of equilibrium between cooling and diffuse heating,
giving an effectively polytropic behavior $\Peq \propto
\rho^{\gamef}$, where $\rho$ is the gas density (see 
Papers II and III for details).\footnote{Note that by ``polytropic''
here we only mean that 
an equation of ``state'' of the form $P \propto \rho^{\gamef}$ is
satisfied, with no implication whatsoever of equilibrium hydrostatic
states, as is the case of equilibrium polytropes in stellar
theory.} Even though the heating and cooling 
functions used do not give a thermally unstable (e.g.,
Field, Goldsmith \& Habing 1969; Balbus 1995)
regime at the temperatures reached by the simulations, they manage to
produce values of $\gamef$ smaller than unity for temperatures in the
range 100 K $< T < 10^5$ K, implying that {\it denser regions are
cooler}. Upon the production of turbulent density
fluctuations, the flow reaches a temperature distribution similar to
that resulting from isobaric thermal instabilities (Field et al.\ 1969),
but without the need for them. Note, however,
that in this case there are no sharp phase transitions. Furthermore,
it is possible that the thermal instability does not have time to form
structures in the presence of supersonic turbulence, since the isobaric mode of
thermal instability in 
a region of size $L$ condenses on characteristic timescale $L/c$, where $c$ is
the sound speed, while the turbulence shears the forming condensations on
timescale $L/v$, where $v>c$ is the characteristic turbulent speed at scale $L$.
Work is in progress to decide on this issue.

\subsection{Cloud Formation}\label{formation}

In the simulations, the largest cloud complexes (several hundred pc)
form simply by gravitational instability. Although in Paper I it was
reported that no gravitationally bound structures were formed, this
conclusion did not take into account the effective reduction of the
Jeans length due to the small $\gamef$ of the fluid. Once this effect
is considered, it is found that the largest scales in the simulations
are unstable. Nevertheless, inside such large-scale clouds, an extrememly
complicated morphology is seen in the higher-density material, as a
consequence of the turbulence generated by the star
formation activity. The medium- and small-scale clouds are
thus turbulent density fluctuations (see also \S\ \ref{corr_vir}).

\subsection{Cloud and magnetic field topology}\label{topology}

\begin{figure} 
%\plotone{fig1.eps}
\plotfiddle{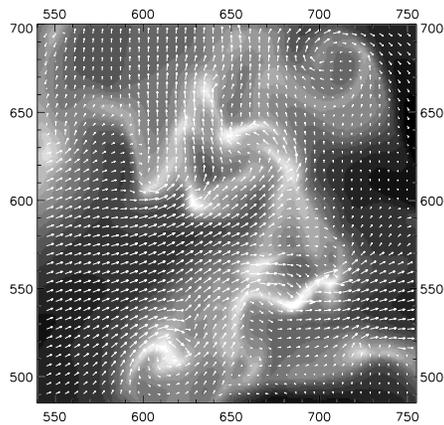}{2in}{0}{30}{30}{-100}{-25}
\caption{Gray-scale image of the logarithm of the density field,
with superimposed magnetic field vectors. Shown is a subfield of $250
\times 250$ pc ($200 \times 200$ pixels), from a simulation at
resolution of 800 grid points per dimension ({VBR97}). The minimum and
maximum magnetic field intensities are 0.12 and 20.1 $\mu$G,
respectively. The axis labels show pixel number. See text for feature
description.}
\label{dens_mag}
\end{figure}

The topology of the clouds formed as turbulent fluctuations in the
simulations is extrememly filamentary. This property apparently
persists in 3D simulations (e.g., Padoan \& Nordlund
1998). Interestingly, the magnetic field also exhibits a morphology 
indicative of significant distortion by the turbulent motions (Paper
II; \S\ \ref{turb_clouds}). The field has a tendency to be aligned with density
features, as shown in fig.\ \ref{dens_mag}. Even in the presence of 
a uniform mean field, motions along the latter amplify the perpendicular
fluctuations due to flux freezing, while at the same time they produce
density fluctuations elongated perpendicular to the direction of
compression. This mechanism also causes many of the density
features to contain 
magnetic field reversals (e.g., the feature with coordinates $x \sim
610$, $y \sim 510$) and bendings (e.g., the feature 
at $x \sim 700, y \sim 550$). It happens also that
magnetic fields can traverse the clouds 
without much perturbation, as seen for example in 
the feature at $x=630, y=600$. These results are consistent with the
observational result that the magnetic field does not seem to vary
much along clouds (Goodman et al.\ 1990), and in general does not
present a unique kind of alignment with the density features. On the
other hand, recent observations have found field bendings and
reversals similar to those described here (Heiles 1997).

It is important to note that the ``pushing'' of the turbulence
on the magnetic field occurs for realistic
values of the energy injection from stars 
and of the magnetic field strength, which
ranges from $\sim 5 \times 10^{-3} \mu$G (occurring at the low density
intercloud 
medium) to a maximum of $\sim 25 \mu$G, which occurs in one of the
high density peaks, although with no unique $\rho$-$B$ correlation
(Paper II). 
%Observationally, larger values of the field occur
%only on much smaller scales than those resolved by our simulations
%(1.25 pc in at resolution of $800^2$ grid points) (Heiles et al.\ 1993). 
Thus, the 
simulations suggest that the effect of the magnetic field is not as
strongly dominating as often assumed in the literature.
This is also in agreement with
the fact that the magnetic and kinetic energies in the simulations are
in near global equipartition at all scales, as shown by their energy
spectra (fig.\ 5 in Paper II), although strong local spatial
fluctuations occur (Paper II; Padoan \& Nordlund 1998).

Finally, note that the fact that the magnetic spectrum
exhibits a clear self-similar (power-law) range, together with the
fact that the fluctuating component of the field is in general
comparable or larger than the uniform field, suggests strongly that
the medium is in a state of fully developed MHD turbulence, rather than
being a superposition of weakly nonlinear MHD waves.

\subsection{Cloud scaling properties}\label{scaling}

An important question concerning the clouds formed in the simulations
is whether they reproduce some well-known observational scaling and
statistical properties of interstellar clouds, most notably the
so-called Larson's relations between velocity dispersion $\Delta v$,
mean density $\rho$ and size $R$ (Larson 1981), and the cloud mass
spectra (e.g., Blitz 1991). \VS, \BP\ \& Rodr\'iguez (1997,
hereafter VBR97) have
studied the scaling properties of the clouds in 
the simulations, finding that the cloud ensemble exhibits a relation
$\Delta v \propto R^{0.4 \pm 0.08}$ and a cloud mass spectrum $dN(M)/dM
\propto M^{-1.44 \pm 0.1}$, both being consistent with observational
surveys, especially those specifically including gravitationally
unbound objects (e.g., Falgarone, Puget \& P\'erault 1992). However,
it was found that 
no density-size relation like that of Larson ($\rho \propto R^{-1}$)
is satisfied by the clouds in the simulations. Instead, the clouds
occupy a triangular region in a $\log \rho$--$\log R$ diagram, as
shown in fig.\ 8 of VBR97, with only its upper envelope being close
to
Larson's relation. This implies the
existence of clouds of very low column density, which can be easily
missed by observational surveys 
if they do not integrate for long enough times. A few observational
works, however, point towards the existence of transients
(Loren 1989; Magnani, la Rosa \& Shore 1993) and low-column density
clouds, with masses much smaller than those estimated from virial
equilibrium, and which exhibit a similar trend to that of the
simulations, namely that of filling an area in a $\log \rho$-$R$ plot,
bounded at large column densities by Larson's density-size relation
(Falgarone et al.\ 1992). 

\section{Tests and Implications of Clouds as Turbulent Density
Fluctuations} \label{turb_clouds}

One crucial implication of the interpretation that clouds are the
turbulent density fluctuations in the ISM is that they are highly
dynamical entities, a conclusion which conflicts with many models
of cloud evolution based on equilibrium configurations, be it by
external pressure confinement (e.g., Maloney 1988; Bertoldi \& McKee
1992; McKee \& Zweibel 1992, hereafter MZ92), or hydrostatic balance between
(microscopic) internal MHD turbulence or wave support and self-gravity (e.g.,
Shu, Adams \& Lizano 1987; Mouschovias 1987; Myers \& Goodman
1988a,b), or combinations thereof (e.g., Zweibel 1990). It is thus
important to turn to 
numerical simulations in which clouds form spontaneously, reproducing
a number of observational cloud properties (\S\S\ \ref{scaling},
\ref{comp_obs}), and investigate
their detailed density, velocity and magnetic fields in order to
decide whether they are in a state of stationary equilibrium, and if
not, whether a dynamical view is still consistent with available
observational data. 

In a recent paper (Ballesteros-Paredes, V\'azquez-Semadeni \& Scalo
1998, hereafter BVS98), we have started to perform work in this
direction. In this section we will review some results from this work
concerning 
the correlation between the density, velocity and magnetic fields, the
evaluation of surface and volume kinetic terms in the Virial Theorem (VT),
some comparisons with observational data, and then some implications,
in particular whether it is feasible to expect the formation of
hydrostatic structures within a turbulent medium.

\subsection{Correlation between variables, and kinetic Virial terms}
\label{corr_vir}

In the scenario that clouds are turbulent density fluctuations, the
essential concept is that they are {\it transient}, i.e.,
non-stationary. Note, however, that such transient character does not
necessarily imply that the clouds will eventually disperse, since,
under appropriate cooling (e.g., Hunter 1979; Hunter et al.\ 1986;
Tohline, Bodenheimer \& 
Christodolou 1987; Paper III; \S\ \ref{hydrostatic?}) they can 
become gravitationally bound and collapse. \BP\ \& \VS\ (1997) have
indeed found in a sample of clouds that a small fraction of them has
gravity strong enough to overwhelm the rest of the terms in the VT. In
this scenario, then, 
clouds form at the collision sites of approaching gas streams, sites
which can contain shocks if the motions are super-magnetosonic, and in
which the highest densities (peaks) are located. We thus expect that, away from
the density maxima, the velocity field be non-zero, and with a net
convergence towards the peak. 

Here it is necessary to note that
cloud-defining algorithms are usually quite arbitrary in the way they
define the cloud's extent, in the sense that they normally search for
a density peak and then extend the cloud out to an arbitrary value,
like the half-maximum (e.g., Williams, de Geus \& Blitz 1994), or
simply threshold the density at some arbitrary value and defining
clouds as connected sets of pixels above the threshold (e.g.,
Ballesteros-Paredes \& V\'azquez-Semadeni 1997). In neither case are
the outer boundaries of the clouds defined on any physical
grounds. Observationally, molecular clouds are often claimed to have
well-defined boundaries (e.g., Blitz 1991), but on the one hand this
is likely an effect of the 
transition to atomic gas in the cloud periphery, and, on the other
hand, there are counter-examples in which no sharp boundaries are
observed (see BVS98 and references therein).

\begin{figure}
\plottwo{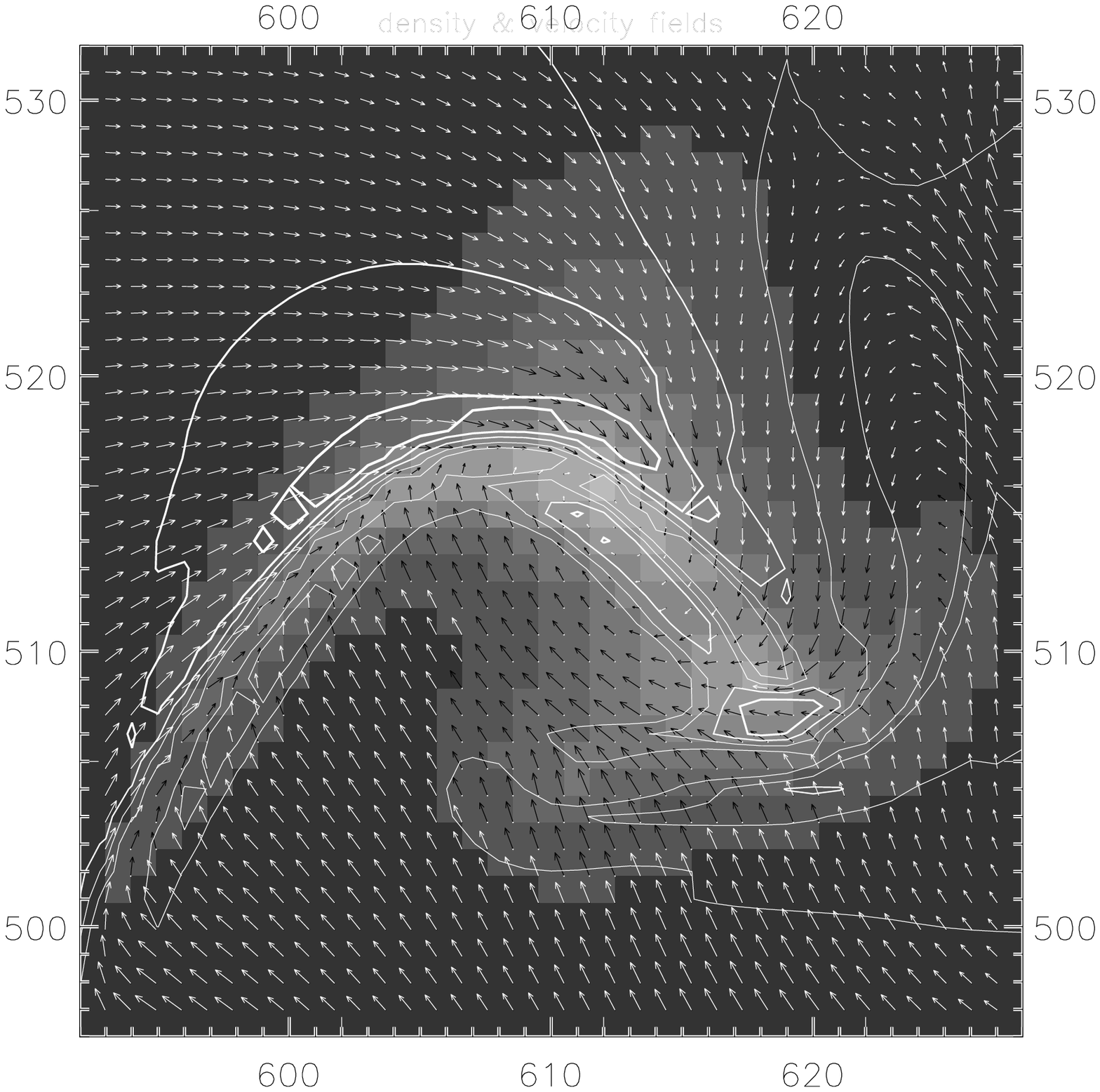}{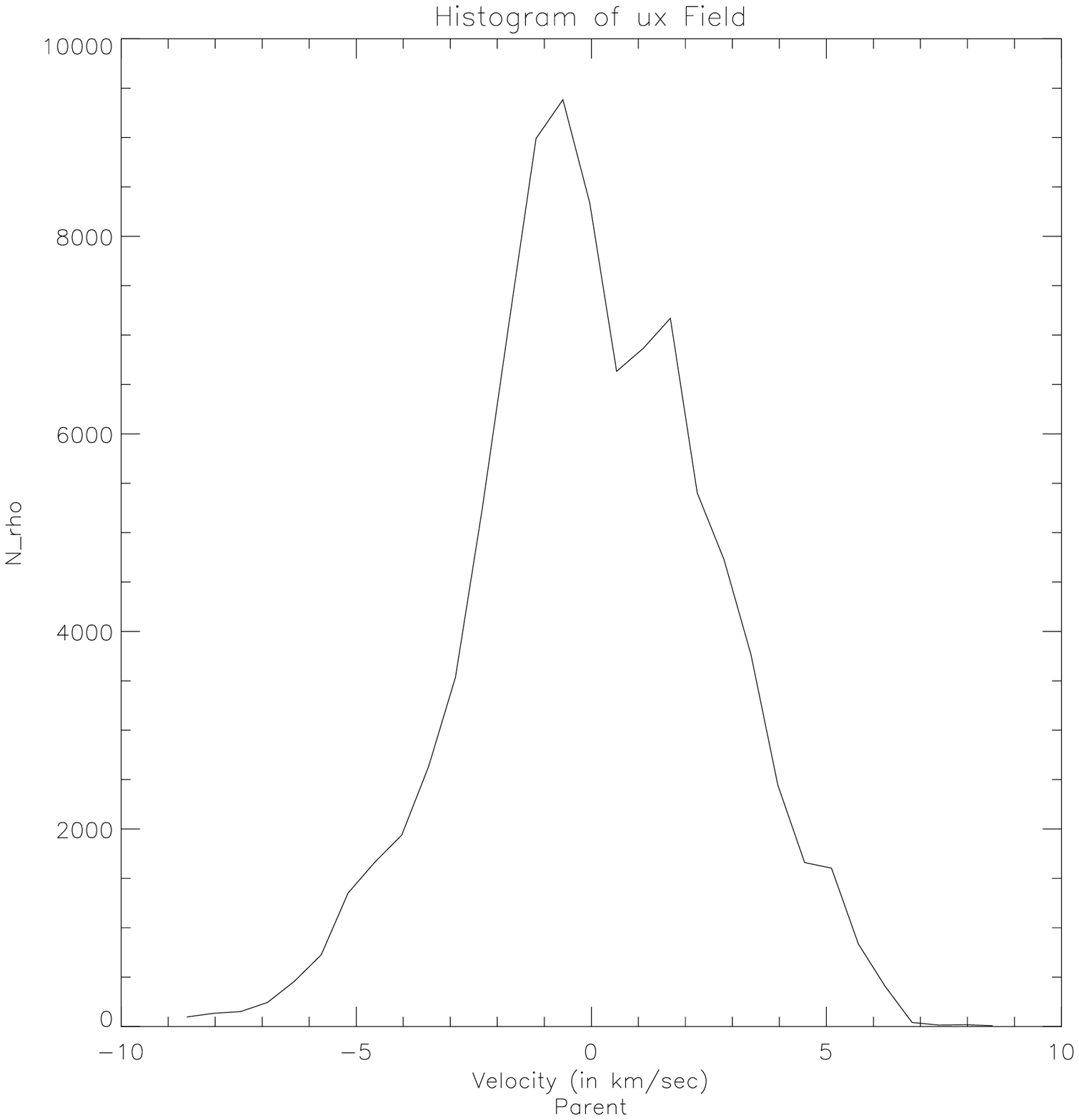}
\caption{{\it a}) Enlargement of the feature at position $x \sim 610$,
$y \sim 510$ in fig.\ \ref{dens_mag}, showing the logarithm of the
density field (gray-scale), velocity field vectors (arrows, black for
super-Alfv\'enic, white for sub-Alfv\'enic) and the magnitude of the
magnetic field (contours, bolder indicating stronger fields). The
cloud boundary is set at a density of 8 cm$^{-3}$m. The maximum
density within the cloud is 55 cm$^{-3}$. {\it b}) 
Density-weighted velocity histogram for the cloud in fig.\
\ref{dens_mag}. Note the multi-component structure.
}
\label{dens_vel_B}
\end{figure}

In view of the above, one expects that the velocity field should be
non-zero when measured along arbitrarily-drawn cloud boundaries. Fig.\
\ref{dens_vel_B}a shows a magnification of the feature at position $x
\sim 610$, $y \sim 510$ in fig.\ \ref{dens_vel_B}. This cloud spans
roughly $40 \times 
30$ pixels in the simulation, (roughly $50 \times 40$ pc). The 
gray-scale denotes the logarithm of the density; the arrows, the
velocity field; and the contours, the magnetic field strength. The
contours go from 4 to 10 $\mu$G in intervals of $1.5 \mu$G, with
thicker contours denoting larger values of $B$. The
black (white) arrows furthermore denote super- (sub-)Alfv\'enic
velocities. Several 
points can be noticed. First, no velocity discontinuity is observed at
the cloud boundaries, nor anywhere inside the cloud, except at its
very center, where the velocity field vanishes. An oblique shock is
seen to delineate the cloud ``tail'' on the left hand side of the
figure. Significant vorticity is also seen around the cloud's
center. All of this is 
consistent with the cloud having been formed by a complex collision of
turbulent streams.

An important consequence of the finiteness and continuity of the
velocity field across the clouds' boundaries is that it implies
significant exchange of mass, momentum and energy through the
boundary. This can be best evaluated by means of the kinetic terms in
the Eulerian Virial Theorem (EVT; see, e.g., Parker 1979; MZ92),
i.e., computed in an Eulerian frame, 
of fixed shape. Note, however, that we allow the Eulerian frame to
move with the mean mass-weighted velocity of the cloud, in order to
eliminate flux through the cloud's boundary due to the cloud's bulk
motion. MZ92 have shown that, in its Eulerian form, the VT
acquires a highly symmetrical form, in that the kinetic contribution
to the VT is of the form $\ekin - \xruu$, where $\ekin=1/2\int_V \rho
u^2 dV$ is a volume-integrated term giving the total kinetic energy in
the cloud's volume $V$, and $\xruu=1/2\oint_S{x_i \rho u_i u_j \hat n_j}
dS$ is a surface integral over the cloud's surface $S$. The latter
term can be interpreted (BVS98) as the sum of the ram pressure plus
the kinetic stresses 
dotted with the normal to the boundary, integrated over the boundary,
in analogy with the corresponding terms for the thermal pressure and
the divergence of the Maxwell stress tensor. Note, however, that, in
contrast with the thermal pressure term, $\xruu$ contains non-isotropic
contributions (BVS98). In the more
familiar Lagrangian form of the VT (see, e.g., Shu 1992) the surface
term does not appear, because the surface moves along with flow,
implying zero flux across it.

BVS98 have calculated the two terms $\ekin$ and $\xruu$ for a
sample of clouds in the simulation mentioned above, finding that in
general both terms are comparable,
evidencing the importance of the surface term compared with the total
kinetic energy. Note that if both terms are equal, they cancel out,
giving no net contribution to the VT, as is well known to be the case
with the corresponding thermal pressure terms (e.g., Shu 1992). On the
other hand, note that there are a few clouds for which $\xruu$ is
negative, meaning that its contribution is of the same sign as that of
$\ekin$, towards an {\it expansion} of the cloud. More generally, the
combined term $\ekin - \xruu$ is in general nonzero, and of either
sign, meaning it may contribute to compressions, expansions or generic
distortions of the clouds.

\subsection{Can hydrostatic structures form in a turbulent medium?}
\label{hydrostatic?}

As mentioned above, a significant number of models of self-gravitating
cloud cores have assumed that these
structures are in (quasi-)hydrostatic equilibrium between self-gravity
and some form of internal support, normally thought to be provided by
a uniform field and Alfv\'en waves (e.g., Shu et al.\ 1987). However,
a serious question is whether such hydrostatic configurations can be
expected to form spontaneously in the turbulent environment in which
they are known to dwell (turbulent molecular clouds). This problem has
been addressed by BVS98 by considering the evolution of turbulently
compressed polytropic masses of gas. Note that in the following
analysis, the internal pressure may be generalized to include sources
of pressure other than thermal, such as magnetic or
microturbulent. Note also that MHD wave pressure is expected to behave
isotropically (Shu et al.\ 1987; McKee \& Zweibel 1995).

It is well known (Chandrasekhar 1961) that in order for a collapsing
spherical mass of gas to eventually stop and reach a hydrostatic
equilibrium, it is necessary that $\gamef > 4/3$, so that the magnitude
of the internal
energy can increase faster than the gravitational energy during the
collapse. This result has been generalized to collapse induced by
compressions in $n \le 3$ dimensions (McKee et al.\ 1993; Paper III),
finding that in this case the critical $\gamef$ for collapse to be
halted is $\gamma_{\rm cr} = 2 (1 -1/n)$. If $\gamef<\gamma_{\rm cr}$,
the $n$-dimensional collapse cannot be halted. Conversely, if
$\gamef > \gamma_{\rm cr}$ an initially gravitationally stable region cannot be
rendered unstable by compression. 

Therefore, BVS98 have pointed out that the formation of hydrostatic
structures requires a {\it change} in $\gamef$ during the collapse
process. This is because, in order to start the collapse of an
initially stable region by a turbulent compression, $\gamef <
\gamma_{\rm cr}$ is required, but in order for internal pressure to
later be able to halt the collapse, it is necessary that $\gamef >
\gamma_{\rm cr}$. Thus, unless $\gamef$ changes during the collapse
process, once the collapse is initiated it cannot be stopped by the
internal pressure. 

\subsection{Implications for the Pressure of Interstellar Gas}
\label{pres_conf} 

The results described in \S\ \ref{corr_vir} have direct consequences
on the underlying hypotheses of standard models of cloud confinement.  
Strictly speaking, clouds in the simulations are not
``confined'', since they are transients. In other words, the left-hand
side of the VT, the second time derivative of the cloud's moment of
inertia ($\ddot I/2$), is in general nonzero (\BP\ \& \VS\
1997). In such a dynamical situation, clouds are {\it not} expected to
be in thermal pressure balance with their surroundings, since the relevant
driving mechanism is the {\it total} pressure, including thermal,
kinetic and magnetic contributions (and cosmic rays, in the actual
ISM). For example, the thermal pressure internal to a cloud may be
increased due to the external ram pressure. 

However, note that,
because the atomic and ionized gas behaves as a 
polytrope with $\gamef < 1$, then the pressure
difference between clouds and their surroundings is rather small. It
is only when molecular densities are reached that $\gamef$ becomes
closer to (or possibly larger than) unity (Scalo et al.\ 1998), and
higher thermal pressure contrasts between clouds and their environment
are observed. In other words, in a turbulent medium in
the presence of 
heating and cooling giving $0 < \gamef < 1$, the near constancy of the
pressure is an incidental consequence of the combination of turbulent
density fluctuation production and the low value of $\gamef$, rather
than a controlling mechanism for cloud confinement. Conversely, note that,
even if a cloud is in thermal pressure balance with its surroundings,
this does not imply that it will be in a static equilibrium
configuration, since ram pressure or simply inertial motions may
distort a cloud.

Molecular clouds, on the other hand, are known to have much higher
pressures, and traditionally this has been attributed to their being
self-gravitating (e.g., Jura 1987). However,
in the present scenario we expect there to also be a significant
contribution due to the conversion of external ram pressure into
thermal pressure at $\gamef \sim 1$. 

The possibility of cloud confinement by turbulent pressure has also
been considered by MZ92. However, note that this mechanism implicitly
assumes that the turbulence is microscopic, i.e., that the turbulent
scales are much smaller than the size of the cloud. Otherwise,
larger-scale modes imply the distortion of Lagrangian cloud boundaries
(which move with the flow), or, equivalently, flux across fixed
Eulerian boundaries.

\subsection{Comparison with Observations} \label{comp_obs}

One crucial test for the dynamical scenario presented here is whether
it compares favorably with observational data. In this section we
discuss several lines of evidence that it does.

A first test is to obtain density-weighted velocity spectra, which can
be compared with observational optically-thin line profiles. Figure
\ref{dens_vel_B}b shows the density-weighted histogram of the
$x$-component of the 
velocity ($u_x$) for the cloud of fig.\
\ref{dens_mag}. This histogram may be compared
with the $^{13}$CO line profile for the Rosette Molecular Cloud (RMC)
shown in fig.\ 4 of Williams, Blitz \& Stark (1995, hereafter
WBS95).\footnote{Although fig.\ 4 in WBS95 shows $^{12}$CO
spectra, which are optically thick, there is little, if any,
qualitative difference with the optically thin $^{13}$CO spectra shown
in fig.\ 5 of that paper.} Both clouds are comparable in many respects:
while the RMC's dimensions (as 
deduced from fig.\ 17 in WBS95) are $\sim 90 \times 70$ pc, the
simulated cloud is $\sim 250 \times 120$ pc. Furthermore, the latter
has a mean density of $\sim 10$ cm$^{-3}$,
while the gas sampled in the spectrum of fig.\ 4 of WBS95 has a mean
density $\sim 15$ cm$^{-3}$.

Comparing the histograms for both of these clouds, we note several
points in common. First, both sets of data have FWHMs of roughly 6
km~s$^{-1}$ when only the main features are considered. Second, both sets
exhibit high-velocity bumps, at several km~s$^{-1}$ from the centroid.
Finally, and probably most importantly, the ``main'' features are seen
to contain substructure in both sets of plots. However, while such
features have been traditionally interpreted as ``clumps'', in our
simulations they originate from extended regions within the cloud.

A second comparison can be done at the level of the cloud lifetimes,
$\tau=L/\Delta v$, where $L$ is the cloud size and  $\Delta v$ its velocity
dispersion.  As an example,
we calculate this for the cloud of fig.\ \ref{dens_vel_B}a. We
find $\Delta v=2.3$ km~s$^{-1}$, and $l\sim 30$ pixels $= 37.5$ pc. Thus,
$\tau \sim 1.6 \times 10^7$ yr. This value is consistent with various
estimates of GMC lifetimes of order a few $\times 10^7$ yr (e.g.,
Bash, Green \&\ Peters 1977; Blitz \& Shu 1980; Larson 1981; Blitz
1994).

A third line of comparison concerns the topology of the density and
magnetic fields. As discussed in \S\ \ref{topology}, the simulation
results appear to be consistent with observations.

\subsection{Applicability to molecular gas} \label{appl_molec}

The ISM simulations discussed in Paper II and BVS98 employ cooling
functions which correspond to atomic and ionized gas,
molecular cooling not being represented. Thus a natural concern is
whether our results are applicable to molecular gas. Although the
obvious ultimate test will be provided by simulations with appropriate
molecular cooling, which will be presented elsewhere, there
are strong reasons to believe the essential dynamical behavior will not be
significantly modified. This is because the basic mechanism is the
production of density fluctuations by turbulent compressive motions,
which is independent of the ``hardness'' of the flow (determined by
$\gamef$). Instead, what appears to change with $\gamef$ is the
amplitude, profile, and 
statistical distribution of the density fluctuations (Passot \&
\VS\ 1998; Scalo et al.\ 1998), but the fluctuations themselves are
necessarily produced by 
negative values of the velocity divergence for whatever value of
$\gamef$. Therefore, all the results discussed above, which are a
consequence of this density fluctuation production mechanism, are
expected to be 
valid independently of the equation of state. In fact, simulations of
isothermal compressible MHD turbulence (e.g., 
Padoan \& Nordlund 1998; Ostriker et al.\ 1998; Mac Low et al.\ 1998),
which may be thought of as using an equation of state more appropriate to
molecular gas,
necessarily exhibit turbulent density fluctuation formation as well.

The effect of self-gravity is likely to slow down
the re-expansion of the denser structures, increasing the lifetime of
structures that do not make it over the ``internal energy barrier''
and collapse, but again the principle that these are dynamical
entities does not seem to be modified.

\section{Conclusions} \label{conclusions}

In this paper we have reviewed the process of interstellar cloud
formation by large-scale turbulence, and some of its consequences
regarding the structure of the resulting clouds found in
high-resolution 2D simulations of the turbulent ISM, such as the effective
polytropic
behavior, the topology and correlation of the density, velocity and
magnetic fields, and the scaling relations that arise. Special
attention was given to the implications of a dynamical conception of
clouds, such as the important mass, energy and momentum fluxes across
Eulerian (fixed) cloud boundaries, which equivalently imply severe
distortions of Lagrangain (moving with the flow) boundaries.

We also discussed the implications on cloud pressure, which we
suggested to be more an incidental consequence of the density field
sculpted by the velocity field in the presence of specific heating and
cooling laws determining the temperature and pressure, than a driving
or confining agent for the clouds. This suggestion should be
applicable at least for all structures in which thermal pressure is
subdominant compared to turbulent and magnetic pressures, i.e., larger
than a few tenths of a pc.

In this regard, we also discussed the applicability of the results,
which are based on simulations that employ cooling functions appropriate
for atomic and ionized gas, to molecular gas. We stressed that the
basic mechanism of turbulent density fluctuation production is
independent of the effective polytropic exponent. In fact, it is
independent of dimensionality as well.

Finally, we discussed the possibility of forming hydrostatic
structures within a turbulent environment in a polytropic fluid, as
would be the case of quiescent molecular cloud cores. We suggested
that this process would require that the effective polytropic exponent
change during an ongoing collapse triggered by the turbulence, a
phenomenon which does not appear likely upon consideration of thermal,
turbulent and magnetic wave pressures. Thus, we suggest that
hydrostatic cores may not exist.

We wish to conclude, however, by noting that the dynamical scenario
outlined here agrees in many respects with observations, as has been
shown from velocity histograms, 
topology of the fields, cloud lifetimes estimates, scaling relations
--except for a density-size relation--, etc. However, it does seem to
differ significantly from stationary, equilibrium models of clouds.

\acknowledgments

I would like to thank Javier Ballesteros-Paredes and John Scalo for a careful
reading of the manuscript and useful comments; J.B.-P.\ for
extensive help with figure preparation, and the Conference
organizers for their 
financial support and hospitality. The simulations were
performed on the CRAY Y-MP 4/64 of DGSCA, UNAM. This work has received partial
financial support from grants DGAPA/UNAM 105295 and CRAY/UNAM
SC-008397.

\end{document}